\documentclass[reprint,
nofootinbib,
amssymb,
aps,
prd,
unsortedaddress,
]{revtex4-1}

\usepackage[intlimits]{amsmath}
\usepackage{accents}

\usepackage[utf8]{inputenc}

\usepackage{graphicx}
\usepackage{dcolumn}
\usepackage{bm}
\usepackage[english]{babel}

\usepackage{tikz}
\usetikzlibrary{arrows.meta}
\usepackage[compat=1.1.0]{tikz-feynman}

\usepackage{verbatim}

\usepackage[colorlinks=true, linkcolor = blue,
urlcolor  = blue,
citecolor = blue]{hyperref}

\begin{document}

\title{A new take on the inflationary quintessence}

\author{Zurab Kepuladze$^{1,2}$ and Michael~Maziashvili$^3$ }
\email{maziashvili@iliauni.edu.ge }

\affiliation{\vspace{0.2cm} $^1$Andronikashvili Institute of Physics, 0177 Tbilisi, Georgia \\ $^2$Institute of Theoretical Physics, Ilia State University, 3/5 Cholokashvili Ave., Tbilisi 0162, Georgia \\  $^3$School of Natural Sciences and Medicine, Ilia State University, 3/5 Cholokashvili Ave., Tbilisi 0162, Georgia}


\begin{abstract} The quintessence field coupled to the cosmic neutrino background (CNB) has been widely	discussed as an alternative mechanism to address the coincidence problem. As it is well known, it is possible to extend such models to obtain quintessential inflation, that is, to incorporate inflationary stage as well. Taking an alternative route, one can start from the well established inflationary models and obtain successful quintessence models at the expense of coupling with the CNB. To Follow this route, we use a slightly reformulated model addressed in PRD {\bf 95}, 123521 (2017). This particular model assumes $\mathcal{Z}_2$ symmetry for both scalar field potential and coupling term, which then breaks down in course of the cosmological evolution. For our discussion, however, the $\mathcal{Z}_2$ symmetry of the potential is not mandatory the model to work. The conventional mechanism of particle production by the oscillating inflaton field (and their subsequent thermalization) remains operative. It is plain to see that the proposed construction can be easily applied for many successful models of inflation to incorporate dark energy at the expense of coupling with the CNB. We address the issue of neutrino nuggets from the quantum field theory point of view. Namely, these nuggets are considered as bound states caused basically by the Yukawa force, which arises in the framework of linear perturbation theory due to exchange of virtual quanta of quintessence field between the neutrinos.   


\end{abstract}

\pacs{Valid PACS appear here}
\maketitle

\section{Preface}

One of the principle motivations for quintessence models of dark energy, introduced in the late 1980's \cite{Weiss:1987xa, Wetterich:1987fm, Peebles:1987ek, Ratra:1987rm}, is to address the cosmic coincidence problem \cite{Zlatev:1998tr, Steinhardt:1999nw}. This problem has two aspects indeed. One of them is to explain the smallness of the present dark energy density and the other one is to figure out what caused the dark energy to activate in the present epoch. A particular class of quintessence models, referred to as trackers, avoid the problem of fine tuning the initial conditions of the scalar field in order to obtain the desired energy density and equation of state at the present time. It is achieved at the expense of introduction of a small scale in the potential - the origin of which may indeed be explained \cite{Binetruy:1998rz, Brax:1999yv}. Another class of models, which we are going to discuss throughout of this paper, explain the coincidence problem by considering a coupling of the quintessence with the cosmic neutrino background (CNB) \cite{Fardon:2003eh, Peccei:2004sz, Wetterich:2007kr, Amendola:2007yx, Brookfield:2005bz}. Roughly speaking, in such models, the idea is to use neutrino mass scale for explaining the smallness of the present dark energy density and the present time activation of quintessence is caused by the back reaction of CNB after neutrinos become non-relativistic. The models of this kind may be used conveniently to unify quintessence with inflation as we have more freedom in choosing the potential. However, in contrast to the quintessential inflation \cite{Geng:2015fla, Ahmad:2017itq, Geng:2017mic}, which aims at the construction of successful inflationary scenario with the "existing" quintessence model, we favor the idea of inflationary quintessence, that is, to construct a successful quintessence model with the use of "good" inflationary models. To be more precise, under the good inflationary models we understand those having a plateau, which provides a slow-roll regime, and a minimum at $\phi=0$ around of which the field starts to oscillate after it exits the slow-roll regime \cite{Linde:2018hmx, Kallosh:2019hzo}. Such models may be considered as main targets for the near-future observational missions. The point is that the upcoming CMB experiments may measure the primordial gravitational wave power spectrum and its amplitude in terms of the tensor to scalar ratio with the precision $5\times 10^{-4} (5\sigma)$ and also aim to improve constraints on the primordial curvature perturbation power spectrum and its tilt \cite{Hanany:2019lle}.   

  The basic idea for constructing such inflationary quintessence models is to use the cosmological symmetry breaking mechanism triggered by the coupling of scalar field with the CNB \cite{MohseniSadjadi:2017pne}. The schematic picture looks as follows. Because of this coupling, the equations of motion contain the Spur of CNB stress-energy tensor which kicks up the scalar field trapped in the minimum $\phi=0$ after the end of preheating and enforces it to roll towards one of the degenerate minima leading thereby to the spontaneous breakdown of $\mathcal{Z}_2$ symmetry. As it is shown in section \ref{earlyt}, it happens shortly after the preheating - around the time of thermalization. After the symmetry breaking, the scalar field evolves adiabatically - tracking roughly the minimum of the effective potential. As a result, the scalar field acquires a non-zero energy density but it is set by the neutrino mass scale and is therefore too small to have any effect at earlier times (see Eq.\eqref{adreuliskalveli}). However, at later times, when CNB gets non-relativistic, the compound of scalar field and CNB starts to act as a dark energy because of adiabatic nature of the scalar field evolution. It does not last forever, since in a while the CNB dilutes enough and its back-reaction providing the slow roll regime for the scalar field becomes negligible. After exiting the slow roll regime, the scalar field, which was monotonically approaching the value $\phi=0$, tends "quickly" towards this point leading to the restoration of $\mathcal{Z}_2$ symmetry.

  One of the most subtle issues from the conceptual point of view is the preheating. Indeed, it may look less natural in the sense that the preheating is usually described by using the quantum theory of a free field with a time-dependent effective mass including the coupling of inflaton with the matter field \cite{Kofman:1994rk, Kofman:1997yn}. This type of preheating can work for fermions as well \cite{Greene:1998nh} but if we are applying this formalism immediately to the neutrinos, we have to distinguish between the background-field time dependence and the real dependence of mass on time. For instance, the coupling of inflaton with the neutrinos responsible for mass variation will not cause particle production at all. On the other hand, the very fact that in some cases the coupling contributes to the real masses while in other cases it just provides a time-dependent background may sound quite unnatural. Emphasizing again, the problem is of conceptual nature rather than technical. It is obvious that the instant preheating mechanism \cite{Felder:1998vq} also suffers from this conceptual problem.

In addition, in the framework of the present model, there are scalar field fluctuations that couple to the CNB resulting in the attractive force between the CNB neutrinos. It can be viewed as a Yukawa force mediated by the scalar quanta as long as we restrict ourselves to the linear perturbations. Under certain circumstances, this approximation may work quite well for describing the formation and subsequent growth of the neutrino nuggets.

Throughout of this paper we are using natural units: $c = \hbar = 1$, in which $G_N^{-1/2}=M_P \approx 1.2\times 10^{28}$\,eV, $H_0 =  74$\, km/sec/Mpc $\approx 1.6\times 10^{-33}$\, eV. Also note that all quantities with subscript or superscript zero refer to the present values.

\section{Description of the model}

We assume a spatially flat FLRW universe with metric

\begin{eqnarray}\label{FLRW-metric}
\mathrm{d}s^2 = \mathrm{d}t^2 - a^2(t)\mathrm{d}\mathbf{x}^2 ~, \nonumber 
\end{eqnarray} and consider a minimal model of $\phi$-$\nu$ coupling given by the action functional

\begin{eqnarray}&&
\int\mathrm{d}^4x\,\sqrt{-g} \left( \frac{g^{\alpha\beta}\partial_\alpha\phi\partial_\beta\phi}{2} \,-\, U(\phi) \,-\,  \frac{M_P^2R}{16\pi } \,+ \right. \nonumber \\&& \left. \frac{i}{2}\left[\bar{\psi}_\nu\gamma^\alpha(x)\mathfrak{D}_\alpha\psi_\nu - \big(\mathfrak{D}_\alpha\bar{\psi}_\nu\big)\gamma^\alpha(x)\psi_\nu  \right] - m(\phi)\bar{\psi}_\nu\psi_\nu  \right) ~. \nonumber
\end{eqnarray} As a next step, for building up the model, the field $\psi_\nu$ is quantized and taken at a finite temperature. That is, $\psi_\nu$ describes a Fermi gas at a finite temperature and is understood to stand for the CNB \cite{Fardon:2003eh, Peccei:2004sz, Wetterich:2007kr, Amendola:2007yx}. Then the equations of motion for the $\phi$-$\nu$ model look as follows

\begin{eqnarray}
&& \dot{\rho}_\nu + 3H(\rho_\nu +p_\nu) =  \frac{\mathrm{d}\ln m_\nu}{\mathrm{d}\phi}(\rho_\nu - 3p_\nu)\dot{\phi} ~, \label{continuity} \\ && 
\ddot{\phi} +3H\dot{\phi} +U'(\phi) = -\frac{\mathrm{d}\ln m_\nu}{\mathrm{d}\phi} (\rho_\nu - 3p_\nu) ~, \label{eqofmot} \\ && H^2 = \frac{8\pi}{3M_P^2}(\rho_\nu+\rho_\phi +\rho_r+\rho_m) \label{Friedmann}~.
\end{eqnarray} For the sake of generality, in Eq.\eqref{Friedmann} we have included the matter and radiation components as well. A remarkable characteristic feature of this system is that at early times, that is, at high temperatures (when neutrinos are relativistic) $p_\nu \approx \rho_\nu/3$ and the right hand sides in Eqs.(\ref{continuity}, \ref{eqofmot}) almost vanish. That means that $\phi$ and $\psi$ fields are nearly decoupled and the dynamics of the scalar field is basically driven by the potential $U(\phi)$. However, in the nonrelativistic regime $p_\nu \ll \rho_\nu$ and the equations of motion take the form

\begin{eqnarray}
&&\dot{\rho}_\nu + 3H\rho_\nu  \approx  \frac{\mathrm{d}\ln m_\nu}{\mathrm{d}\phi}\rho_\nu \dot{\phi} ~, \nonumber \\&&
\ddot{\phi} +3H\dot{\phi} +U'(\phi) \approx -\frac{\mathrm{d}\ln m_\nu}{\mathrm{d}\phi} \rho_\nu  ~, \nonumber
\end{eqnarray} or, in terms of the CNB number density\footnote{For CNB is in the non-relativistic regime, $\rho_\nu \approx n_\nu m_\nu$.} $n_\nu \approx \rho_\nu / m_\nu(\phi)$,

\begin{eqnarray}\label{phiindependent}
&&\dot{n}_\nu + 3Hn_\nu  \approx  0 ~, \nonumber   \\\label{autonomous}
&&\ddot{\phi} +3H\dot{\phi} +U'(\phi) \approx -m'_\nu(\phi) n_\nu  ~.
\end{eqnarray} The Eq.\eqref{autonomous} describes the motion of scalar field in the modified potential

\begin{eqnarray}\label{efekturipotentsiali}
U_{eff} = U(\phi)+m_\nu(\phi)n_\nu(t) ~. 
\end{eqnarray} Assume that the effective  potential has a local minimum, $\phi_+$, in which the scalar field is trapped. The presence of $n_\nu(t)$ in the effective potential indicates that the minimum itself is time-dependent. The model to work, the effective potential should provide the slow roll, 

\begin{eqnarray}
	\left|\ddot{\phi_+}\right| \ll H\left|\dot{\phi}_+\right| ~, ~~ \dot{\phi}_+^2 \ll U_{eff}(\phi_+) ~. \nonumber 
\end{eqnarray}

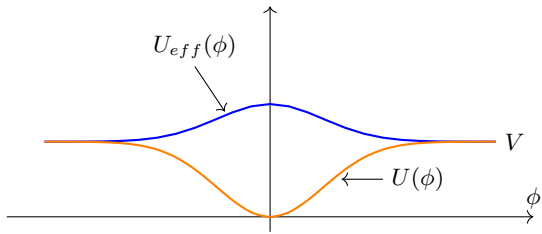
\begin{figure}[h]
	\centering
	\begin{tikzpicture}[domain=-3:3] 
	\draw[->] (0,-0.2) -- (0,2.8) ;
	\draw[->] (-3.5,0) -- (3.5,0) node[above] {$\phi$};
	\draw[color=blue, thick]    plot (\x,{1-exp(-1*\x*\x)+1.5*exp(-1*\x*\x)})   node[right, black] {$V$}; 
	\draw[<-]  (-0.6,1.4) -- (-1,2) node[above] {$U_{eff}(\phi)$} ;
\draw[<-]  (0.95,0.5) -- (1.5,0.5) node[pos=1, anchor= west] {$U(\phi)$} ;
	\draw[color=orange, thick] plot [smooth] (\x,{1-exp(-1*\x*\x)}) ;
	\end{tikzpicture}
	\caption{The potentials for the model discussed in \cite{MohseniSadjadi:2017pne}.} \label{fig1}
\end{figure}

The first model we want to consider is obtained by the reformulation of the one addressed in \cite{MohseniSadjadi:2017pne}. It consists of the $\mathcal{Z}_2$ symmetric potential and $\phi$-$\nu$ coupling of the form

\begin{eqnarray}\label{potential}
U(\phi) = V\left(1-\mathrm{e}^{-\alpha \phi^2/M^2_P}\right) ~, ~ m_\nu(\phi) = \mu_\nu\mathrm{e}^{-\beta \phi^2/M^2_P} ~. ~~~
\end{eqnarray} This model is clearly motivated by the paper \cite{Pietroni:2005pv}. In \cite{MohseniSadjadi:2017pne} it is assumed that $V$ is of the order of the present dark energy density. We find this assumption undesirable because if we take the existence of the present cosmological constant for granted - then there is nothing to explain as its existence will not spoil anything in the past \cite{Weinberg:1987dv} and therefore one may not worry about the hiding of that small cosmological constant in the past \cite{MohseniSadjadi:2017pne}. As to the scenario, it looks as follows. It is assumed that $\phi = 0$ before neutrinos enter a non-relativistic regime and after that $\phi$ is driven by the effective potential \eqref{efekturipotentsiali}. The effective potential turns the point $\phi =0$ into the local maximum, as it is depicted in Fig.\ref{fig1}, and field starts moving either left or right providing the present dark energy $V$.

 After a while, when $n_\nu(t)$ dilutes enough, the potential approaches its initial form and the field, which is rolling back, will reach the point $\phi =0$ restoring thereby the initial $\mathcal{Z}_2$ symmetry. In itself, the mechanism used in this scenario for hiding the dark energy both in the past and in the future looks quite attractive.

In order to unify dark energy and inflation, we reformulate the above model by assuming large $V$ in Eq.\eqref{potential}. This way one obtains a typical example of the T-model \cite{Carrasco:2015rva}. Thus, we consider the effective potential

\begin{eqnarray}\label{natural}
U_{eff}(\phi) = V\left(1-\mathrm{e}^{-\alpha \phi^2/M^2_P}\right) + n_\nu \mu_\nu\mathrm{e}^{-\beta \phi^2/M^2_P} ~, \nonumber
\end{eqnarray} where, the parameter $V$ (which is understood to be large) together with the parameter $\alpha$ is "determined" from the requirements of inflation, while the parameter $\beta$ is set from the requirements of present dark energy.

\section{$\alpha, V$ parameters}
\label{parameters} 

 The potential $U(\phi)$ has an infinitely long plateau for large values of $|\phi|$ starting roughly at $\phi^2 \simeq M_P^2/\alpha$. The potentials with plateau provide perfect conditions for the slow-roll inflation as the field rolling down will arrive at the attractor trajectory from a very wide range of initial conditions and are favorable by the present cosmological data. 
In this section we estimate the inflationary observables for the
model discussed above. The slow-roll parameters are defined as

\begin{eqnarray}
\epsilon = \frac{M_P^2}{16\pi }\left(\frac{U'}{U}\right)^2 ~, ~~ \eta =\frac{M_P^2}{8\pi } \frac{U''}{U} ~, ~~ \xi = \frac{M_P^4}{(8\pi)^2}\frac{U'U'''}{U^2} ~. \nonumber 
\end{eqnarray} The end of inflation occurs for $\phi_f$ at which $\epsilon(\phi_f) \simeq 1$. From this condition one finds  

\begin{eqnarray}&&
\alpha = 1 \Rightarrow	\phi_f \simeq 0.27\times M_P ~,~  \alpha = 4 \Rightarrow	\phi_f \simeq 0.25\times M_P \nonumber \\&& \alpha = 9 \Rightarrow	\phi_f \simeq 0.22\times M_P ~,~ \alpha = 16 \Rightarrow	\phi_f \simeq 0.2\times M_P ~.\nonumber 
\end{eqnarray} The number of $\mathrm{e}$-foldings is

\begin{eqnarray}
N = -\frac{8\pi}{M_P^2}\int_{\phi_i}^{\phi_f}\mathrm{d}\phi \, \frac{U}{U'} = -4\pi \int_{\phi_i}^{\phi_f}\mathrm{d}\phi \, \left[ \frac{\mathrm{e}^{\alpha\phi^2/M_P^2}}{\alpha\phi} - \frac{1}{\alpha\phi} \right] = \nonumber \\  \frac{4\pi}{\alpha} \left[ \frac{\text{Ei}\big(\alpha\phi_i^2/M_P^2\big)}{2} - \frac{\text{Ei}\big(\alpha\phi_f^2/M_P^2\big)}{2} + \ln\left(\frac{\phi_i}{\phi_f}\right)\right]  ~, \nonumber  
\end{eqnarray} which after demanding $N=60$ determines the initial values of the field as

\begin{eqnarray}&&
\alpha = 1 \Rightarrow	\phi_i \simeq 1.35\times M_P ~,~  \alpha = 4 \Rightarrow	\phi_i \simeq 1.1\times M_P \nonumber \\&& \alpha = 9 \Rightarrow	\phi_i \simeq 0.81\times M_P ~,~ \alpha = 16 \Rightarrow	\phi_i \simeq 0.65\times M_P ~.\nonumber 
\end{eqnarray} The slow-roll parameters can be used to express the spectral index, its derivative and tensor-to-scalar ratio as

\begin{eqnarray}
n_s = 1 -6\epsilon + 2\eta ~, ~ \frac{\mathrm{d}n_s}{\mathrm{d}\ln k} = 16\epsilon \eta -24\epsilon^2 - 2\xi ~, ~ r = 16\epsilon  ~. \nonumber 
\end{eqnarray} To fit the present observational data \cite{Akrami:2018odb} 

\begin{eqnarray}
n_s = 0.9649 \pm 0.0042  ~,  ~ r < 0.06 ~, \nonumber \\ \frac{\mathrm{d}n_s}{\mathrm{d}\ln k} =  - 0.0045 \pm 0.0067 ~,~~~~~~ \nonumber 
\end{eqnarray} the parameter $\alpha$ should satisfy $\alpha \gtrsim 6.4$.

As to the the energy scale of inflation, $V$, it is commonly expected to lie approximately between the TeV and Planck scales. It is related to the amplitude of tensor modes 

\begin{eqnarray}
V^{1/4} \simeq 3.3\times 10^{16} r^{1/4}\text{GeV} ~, \nonumber 
\end{eqnarray} indicating that the detectable gravitational waves require $V^{1/4} \simeq 10^{16}$GeV. That is the energy scale considered in \cite{Carrasco:2015rva} but, in general, such a big value is not typical for the existing models of inflation. In what follows we admit the whole "possible" range of parameter $V$ but for the discussion of nuggets it is favorable to take this parameter near the lower bound (see section \ref{nnuggets}).

\section{Onset of dark energy}

\begin{figure}[h]
	\centering
	\begin{tikzpicture}[domain=-3:3] 
	\draw[->] (0,-0.2) -- (0,3) ;
	\draw[->] (-3.5,0) -- (3.5,0) node[above] {$\phi$};
	\draw[color=blue, thick, domain=-3.2:3.2,]    plot [smooth] (\x,{2-2*exp(-0.3*\x*\x)+1*exp(-3*\x*\x)})   ; 
	\draw[color=orange, thick, domain=-3.2:3.2,] plot [smooth] (\x,{2-2*exp(-0.3*\x*\x)}) ;
	\draw[<-]  (-2.1,1.4) -- (-2.5,0.8) node[below] {$U(\phi)$} ;
	\draw[<-]  (-0.4,0.8) -- (-1,1.5) node[above] {$U_{eff}(\phi)$} ;
	\draw [blue, dashed] (0.75,0.5) -- (0.75,0) node[black, below]{$\phi_{+}$}; 
	\draw [blue, dashed] (-0.75,0.5) -- (-0.75,0) node[black, below]{$\phi_{-}$}; 
	\fill[blue] (0,1) circle (0.06cm);
	\draw[<-]  (0.06,1.05) -- (0.8,1.9) node[above] {$~~ n_\nu \mu_\nu$} ;
	\end{tikzpicture}
	\caption{The symmetry breaking effective potential.} \label{fig2}
\end{figure}
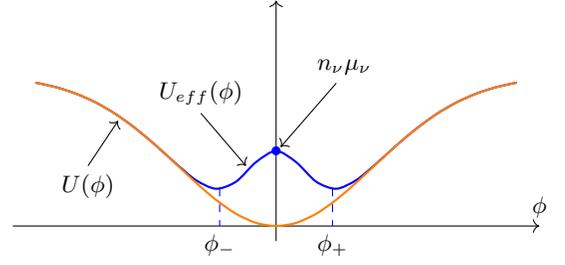

In order to obtain dark energy, the effective potential \eqref{efekturipotentsiali} should provide the slow roll regime. When $\alpha > \beta$, the extremum $\phi = 0$ gives the only minimum. Putting $\alpha < \beta$ and at the same time demanding

\begin{eqnarray}\label{piroba}
\frac{\alpha V}{\beta  n_\nu \mu_\nu} < 1 ~, \nonumber 
\end{eqnarray} we will have two minimum points

\begin{eqnarray}\label{minimumebi}
\frac{\phi_{\pm}}{M_P} = \pm\sqrt{ \frac{1}{(\beta -\alpha)}\ln \frac{\beta  n_\nu \mu_\nu}{\alpha V}} ~,
\end{eqnarray} and one maximum at $\phi=0$ as it is shown in Fig.\ref{fig2}. In this case the symmetry breaking takes place in accordance with the scenarios described in \cite{Pietroni:2005pv, MohseniSadjadi:2017pne}, however, as it is discussed in the following section, it occurs shortly after the preheating. Around this time, the effective potential develops the minimums for which $U(\phi_\pm)\lesssim U(0)$ and field moves to one of them and then follows the dynamics of this minimum. For definiteness let us take this minimum to be $\phi_+$. Returning to the non-relativistic regime of CNB, it is plain to see that the neutrino masses increase in such a way

\begin{eqnarray}
m_\nu(\phi_+) = \mu_\nu \mathrm{e}^{-\beta \phi^2_+/M_P^2} \approx  \frac{\alpha V}{\beta n_\nu} ~, \nonumber 
\end{eqnarray} that the neutrino energy density

\begin{eqnarray}\label{constant}
\rho_\nu = n_\nu m_\nu \approx \frac{\alpha V}{\beta } = \text{const}. ~.  
\end{eqnarray} This kind of behavior of CNB in the non-relativistic regime lasts until the symmetry restoration takes place, which in view of Eq.\eqref{minimumebi} occurs when $n_\nu$ drops down to

\begin{eqnarray}\label{fazurigadasvla}
n_\nu =\frac{\alpha V}{\beta   \mu_\nu}  ~. 
\end{eqnarray} After the symmetry restoration, the mass of neutrino $\mu_\nu$ becomes time independent. As it is discussed below, $\mu_\nu \simeq 8000\times m^0_\nu$, where $m^0_\nu$ stands for the present value of the mass.

Now let us see if the model provides a slow roll regime at present. For this purpose, we shall verify the condition

\begin{eqnarray}
\frac{\dot{\phi}_+^2}{2} \ll U(\phi_+) ~, \nonumber 
\end{eqnarray} which is tantamount to

\begin{eqnarray}\label{slowrolling}
&&\frac{M_P^2 9H^2}{(\beta -\alpha)8\ln \frac{\mu_\nu}{m_\nu(\phi_+)}} \ll  \nonumber \\ && V\left(1 - \exp\left(-\frac{\alpha}{\beta -\alpha}\ln \frac{\mu_\nu}{m_\nu(\phi_+)}\right)\right) ~. 
\end{eqnarray}

 In view of Eq.\eqref{constant}, we need to demand $\beta \gg \alpha$ in order to ensure $\rho^0_\nu < \rho_c$, where $\rho_c \equiv 3H_0^2M_P^2/8\pi$ stands for the critical energy density. Recalling that the parameters $V$ and $\alpha$ are set from having a successful inflationary model,

\begin{eqnarray}
	\alpha \sim 1 ~, ~~ \text{TeV} \lesssim  V^{1/4} \lesssim 10^{13}\text{TeV} ~, \nonumber 
\end{eqnarray} one can make the following order of magnitude estimate

\begin{eqnarray}
	\beta \sim \alpha \frac{ V}{H_0^2M_P^2} \sim 10^{122} \frac{  V}{M_P^4} ~\Rightarrow ~ \beta \gtrsim 10^{58} ~.  \nonumber 
\end{eqnarray} In view of this, the Eq.\eqref{slowrolling} simplifies to

\begin{eqnarray}
	\frac{M_P^2H^2}{V} \ll \alpha  \ln^2\big(\mu_\nu/m_\nu(\phi_+)\big) ~,
\end{eqnarray} and is satisfied with an extremely high accuracy at present if we take $\ln\big(\mu_\nu/m^0_\nu(\phi_+)\big)=9$, which follows from the requirement that the equation-of-state-parameter $= -0.9$. Namely, under assumption of slow roll, the dark energy density is given by \cite{Peccei:2004sz, Franca:2003zg}

\begin{eqnarray}
\rho_{dark} = U(\phi_+) + m_\nu(\phi_+)n_\nu \approx \frac{\alpha V}{\beta}\left( \ln \frac{\mu_\nu}{m_\nu} +1\right) ~, \nonumber 
\end{eqnarray} and correspondingly \begin{eqnarray}
\omega^0 = \frac{p^0_{dark}}{\rho^0_{dark}} \approx -\frac{U(\phi_+)}{U(\phi_+) + \rho_\nu } = -\frac{\ln \mu_\nu /m^0_\nu}{\ln \mu_\nu /m^0_\nu +1} = -0.9 ~, \nonumber 
\end{eqnarray} results in: $\ln\big(\mu_\nu/m^0_\nu(\phi_+)\big)=9$.

Next, we have to tune the parameters $\alpha, \beta, V$ in order to obtain: $\rho^0_{dark} = 0.7 \rho_c$. That is, we must demand

\begin{eqnarray}
\frac{10\alpha V}{\beta} = \frac{0.7\times 3M_P^2H_0^2}{8\pi} ~. \nonumber 
\end{eqnarray} A crude estimate of the duration of present accelerated expansion maybe made by assuming that the expansion has an exponential character. Then, from Eqs.(\ref{constant}, \ref{fazurigadasvla}) one obtains

\begin{eqnarray}
	\mathrm{e}^{-3H_0(t-t_0)} \simeq \mathrm{e}^{-9} ~, \Rightarrow ~ t-t_0 \sim 3H^{-1}_0 ~.  \nonumber 
\end{eqnarray}

It is also important to clarify the question - when does the accelerated expansion start? For this purpose, we have to verify the condition

\begin{eqnarray}
3p_\phi < - \rho_\phi   - 2\rho_r - \rho_m - \rho_\nu ~, \nonumber 
\end{eqnarray} which follows from 

\begin{eqnarray}
\frac{\ddot{a}}{a} = \dot{H}+H^2 = -\frac{4\pi}{3M_P^2}\left( \rho_\phi+3p_\phi +2\rho_r+\rho_m+\rho_\nu \right) ~. \nonumber 
\end{eqnarray} That is, we have to check

\begin{eqnarray}
	\frac{\alpha V}{\beta} \left(1-2\ln\frac{ \mu_\nu (1+z)^3}{m^0_\nu} \right) < ~~~~~~~\nonumber \\ -2\rho_r^0(1+z)^4-\rho_m^0(1+z)^3 ~. \nonumber 
\end{eqnarray} Since we know that $\alpha V/\beta = 0.07\rho_c, \ln  \big(\mu_\nu/m^0_\nu\big) = 9, \Omega_r = 5.38\times 10^{-5}, \Omega_m = 0.31$, this relation can be put in the form     

\begin{eqnarray}
	-0.07 \left[ 17 + 6\ln(1+z) \right] <  -2\times 5.38\times 10^{-5}(1+z)^4 - \nonumber \\ 0.31(1+z)^3 ~~\Rightarrow ~~  z \lesssim 0.65 ~.~~~~~~~~~~~~~~~~~ \nonumber 
\end{eqnarray}

In the next section we go back to the early universe to describe the evolution of $\phi$-$\nu$ mixture at early times.

  \section{Early times}
\label{earlyt}

 To elucidate the model further, it is expedient to proceed the discussion in terms of the CNB temperature by using the phase space distribution function for the free-streaming neutrinos \cite{Peccei:2004sz}

\begin{eqnarray}&&
\rho_\nu = \frac{\mathsf{g}}{a^3}\int\frac{\mathrm{d}^3k}{(2\pi)^3}\, \frac{\varepsilon_\nu(\mathbf{k})}{\mathrm{e}^{k/aT_\nu} + 1} ~,  \nonumber \\&& p_\nu  =  \frac{\mathsf{g}}{3a^5}\int\frac{\mathrm{d}^3k}{(2\pi)^3}\frac{k^2}{\varepsilon_\nu(\mathbf{k})\Big(\mathrm{e}^{k/aT_\nu} + 1\Big)} ~, \nonumber \\ && \varepsilon_\nu(\mathbf{k}) = \sqrt{\frac{\mathbf{k}^2}{a^2}+m_\nu^2} ~,   \nonumber
\end{eqnarray} where $\mathsf{g}$ counts all effectively contributing neutrino degrees of freedom. If we are restricting ourselves to the one species of neutrino, then $\mathsf{g}=4$. Evaluating the time-derivative $\dot{\rho}_\nu$, one obtains

	\begin{eqnarray}\label{tsarmoebuli}&&
	\dot{\rho}_\nu = -\frac{3\mathsf{g}\dot{a}}{a^4}\int\frac{\mathrm{d}^3k}{(2\pi)^3}\frac{\varepsilon_\nu(\mathbf{k})}{\mathrm{e}^{k/aT_\nu} + 1} + \nonumber \\ && \frac{\mathsf{g}}{a^3}\int\frac{\mathrm{d}^3k}{(2\pi)^3}\, \frac{\dot{\varepsilon}_\nu(\mathbf{k})}{\mathrm{e}^{k/aT_\nu} + 1} - \nonumber \\ &&   \frac{\mathsf{g}}{a^3}\int\frac{\mathrm{d}^3k}{(2\pi)^3}\, \frac{\varepsilon_\nu(\mathbf{k})\, \mathrm{e}^{k/aT_\nu}}{\Big(\mathrm{e}^{k/aT_\nu} + 1\Big)^2} \, \frac{\mathrm{d}}{\mathrm{d}t}  \frac{k}{aT_\nu} ~, \nonumber  
	\end{eqnarray} where the equation

	\begin{eqnarray} \frac{\mathrm{d}}{\mathrm{d}t}  \frac{1}{aT_\nu}  = 0 ~~\Rightarrow~~ aT_\nu = \text{const.} ~,  \nonumber 
	\end{eqnarray} determines the temperature as a function of time and the remaining terms, after substituting 
	
	\begin{eqnarray}
	\dot{\varepsilon}_\nu(\mathbf{k}) = \frac{-a^{-3}\dot{a}k^2+m_\nu m_\nu'\dot{\phi}}{\varepsilon_\nu(\mathbf{k})} ~, \nonumber
	\end{eqnarray} result in Eq.\eqref{continuity}

\begin{eqnarray}
&&\dot{\rho}_\nu + 3H(\rho_\nu +p_\nu) =    \frac{\mathrm{d}\ln m_\nu}{\mathrm{d}\phi}\dot{\phi}(\rho_\nu - 3p_\nu) = \nonumber\\ && \frac{\mathsf{g}m_\nu m_\nu'\dot{\phi}}{a^3}\int\frac{\mathrm{d}^3k}{(2\pi)^3}\, \frac{1}{\varepsilon_\nu(\mathbf{k})\Big(\mathrm{e}^{k/aT_\nu} + 1\Big)} ~. \nonumber
\end{eqnarray} Thus, one finds that

\begin{eqnarray} \label{rhosminus3p}
	\rho_\nu - 3p_\nu = \frac{ \mathsf{g} T^4_\nu}{2\pi^2} \, \frac{m^2_\nu}{T^2_\nu}\int_0^\infty\mathrm{d}\xi \, \frac{\xi^2}{\sqrt{\xi^2+m^2_\nu/T^2_\nu}\left(\mathrm{e}^\xi +1\right)} ~. ~~~
\end{eqnarray} In the limit $m_\nu/T_\nu \gg 1$, the expression \eqref{rhosminus3p} simplifies to

\begin{eqnarray}\label{nonrel}
\rho_\nu - 3p_\nu \simeq \frac{ m_\nu\mathsf{g} T^3_\nu}{2\pi^2} \, \int_0^\infty\mathrm{d}\xi \, \frac{\xi^2}{\mathrm{e}^\xi +1} = \frac{ 3\zeta(3) \mathsf{g} m_\nu T^3_\nu}{4\pi^2}   ~, ~~ \nonumber 
\end{eqnarray} while in the case $m_\nu/T_\nu \ll 1$ it can be approximated by 

\begin{eqnarray}\label{ultrarel}
\rho_\nu - 3p_\nu \simeq  \frac{ \mathsf{g} T^4_\nu}{2\pi^2} \, \frac{m^2_\nu}{T^2_\nu} \int_0^\infty\mathrm{d}\xi \, \frac{\xi}{\mathrm{e}^\xi +1} = \frac{ \mathsf{g} m^2_\nu T^2_\nu}{24}  ~, 
\end{eqnarray} It is worth paying attention that the phase-space distribution function for neutrinos what we have used above is valid after the neutrinos are decoupled from the rest of the universe. Roughly, the decoupling temperature is $1$\,MeV - much bigger than $m_\nu \leq \mu_\nu \simeq 8$KeV. Above the decoupling temperature one has to use the equilibrium distribution function 

\begin{eqnarray}&&
\rho_\nu = \frac{\mathsf{g}}{a^3}\int\frac{\mathrm{d}^3k}{(2\pi)^3}\, \frac{\varepsilon_\nu(\mathbf{k})}{\mathrm{e}^{\varepsilon_\nu(\mathbf{k})/T_\nu} + 1} ~,  \nonumber \\&& p_\nu  =  \frac{\mathsf{g}}{3a^5}\int\frac{\mathrm{d}^3k}{(2\pi)^3}\frac{k^2}{\varepsilon_\nu(\mathbf{k})\Big(\mathrm{e}^{\varepsilon_\nu(\mathbf{k})/T_\nu} + 1\Big)} ~, \nonumber \\ && \varepsilon_\nu(\mathbf{k}) = \sqrt{\frac{\mathbf{k}^2}{a^2}+m_\nu^2} ~,   \nonumber
\end{eqnarray} which results in 

\begin{eqnarray}\label{tsonastsoruli}
\rho_\nu - 3p_\nu = \frac{ \mathsf{g} T^4_\nu}{2\pi^2} \, \frac{m^2_\nu}{T^2_\nu}\int_{m_\nu/T_\nu}^\infty\mathrm{d}\xi \, \frac{\sqrt{\xi^2-m^2_\nu/T^2_\nu}}{\mathrm{e}^\xi +1} ~. 
\end{eqnarray} One sees from Eq.\eqref{tsonastsoruli} that in the ultra-relativistic limit we arrive again at the Eq.\eqref{ultrarel}. We shall first consider CNB below the decoupling temperature - that is the free streaming regime. In this case the temperature dependent effective potential can be written as $U(\phi)+\rho_\nu(\phi, T_\nu)$,  

\begin{widetext}
\begin{eqnarray}\label{zogadi}
	U_{eff}(\phi, T_\nu) =   U(\phi) + \frac{ \mathsf{g} T^4_\nu}{2\pi^2} \int_0^\infty\mathrm{d}\xi \, \frac{\xi^2\sqrt{\xi^2+m^2_\nu/T^2_\nu}}{\mathrm{e}^\xi +1} =   V\Big(1-\mathrm{e}^{-\alpha\phi^2/M_P^2}\Big) + \frac{ \mathsf{g} T^4_\nu}{2\pi^2}\int_0^\infty\mathrm{d}\xi \, \frac{\xi^2}{\mathrm{e}^\xi +1} \sqrt{\xi^2+\frac{\mathrm{e}^{-2\beta \phi^2/M_P^2}}{(T_\nu/\mu_\nu)^2}} ~. ~~~
\end{eqnarray}\end{widetext} For our discussion we need the point $\phi=0$ to be a maximum for the present value of temperature $T_\nu \ll \mu_\nu$. For this reason, let us evaluate the second derivative of \eqref{zogadi} at $\phi = 0$

\begin{eqnarray}
	U''_{eff}(\phi=0, T_\nu) =  \frac{2\alpha V}{M_P^2} - \nonumber \\  \frac{\beta \mathsf{g} \big(\mu_\nu T_\nu\big)^2}{\pi^2 M_P^2}\int_0^\infty\mathrm{d}\xi \, \frac{\xi^2}{(\mathrm{e}^\xi +1)\sqrt{\xi^2+\left(\frac{\mu_\nu}{T_\nu}\right)^2}} ~, \nonumber 
\end{eqnarray} one finds that $\phi = 0$ corresponds to maximum if

\begin{eqnarray}\label{minimax}
	\frac{2\alpha \pi^2 V}{\beta\mathsf{g}\big(\mu_\nu T_\nu\big)^2} < \int_0^\infty\mathrm{d}\xi \, \frac{\xi^2}{(\mathrm{e}^\xi +1)\sqrt{\xi^2+\left(\frac{\mu_\nu}{T_\nu}\right)^2}} ~. 
\end{eqnarray} Let us note that if $\phi=0$ represents a maximum for the present value of temperature, then it automatically implies that this point is maximum for higher temperatures as well. Namely, the integral in Eq.\eqref{minimax} increases monotonically to the value $\pi^2/12(\approx 0.82)$ as the temperature goes to infinity while the left-hand side of this inequality becomes decreasing as the temperature increases.

Let us look at the ultra-relativistic regime, which, in view of Eqs.(\ref{ultrarel}, \ref{zogadi}) enables one to put the effective potential in a simple form

\begin{eqnarray}\label{numbered}
		U_{eff} \approx V\left(1-\mathrm{e}^{-\alpha\phi^2/M_P^2}\right) + \frac{7\pi^2\mathsf{g}T^4_\nu}{240} + \nonumber \\  \frac{\mathsf{g}(\mu_\nu T_\nu)^2\mathrm{e}^{-2\beta\phi^2/M_P^2}}{48} ~. 
\end{eqnarray} The minimum points in the ultra-relativistic regime are defined by

\begin{eqnarray}\label{relatminimumebi}
	\frac{\phi_\pm}{M_P} = \pm \sqrt{\frac{1}{2\beta - \alpha}\ln \frac{\beta\mathsf{g}(\mu_\nu T_\nu)^2}{48\alpha V}} ~,
\end{eqnarray} while they look as (see Eq.\eqref{minimumebi})

\begin{eqnarray}
\frac{\phi_\pm}{M_P} = \pm \sqrt{\frac{1}{\beta - \alpha}\ln \frac{3\zeta(3)\beta\mathsf{g}T_\nu^3}{4\pi^2\alpha V}} ~,\nonumber 
\end{eqnarray} in the non-relativistic regime. From Eq.\eqref{relatminimumebi} one obtains that the neutrino mass in the early universe is much smaller than the present value

\begin{eqnarray}\label{masa}
	m_\nu(\phi_+) = \sqrt{\frac{48\rho^0_\nu}{\mathsf{g}}} \, \frac{1}{T} ~. \nonumber 
\end{eqnarray} Apart from this, the energy density of the scalar field 

\begin{eqnarray}\label{adreuliskalveli}
	&&\frac{\dot{\phi}^2_+}{2} + U(\phi_+) \approx \nonumber \\&&  \frac{M_P^2H^2}{2(2\beta -\alpha)\ln \frac{\mathsf{g}(\mu_\nu T_\nu)^2}{48\rho^0_\nu}} + \frac{\rho^0_\nu}{2}\ln \frac{\mathsf{g}(\mu_\nu T_\nu)^2}{48\rho^0_\nu} ~,  
\end{eqnarray} is now negligibly small as compared to the neutrino energy density

  \begin{eqnarray}
  \rho_\nu \approx \frac{7\pi^2\mathsf{g}T_\nu^4}{240}~. \nonumber 
  \end{eqnarray} This conclusion is almost obvious by noting that $M_P^2H^2 \simeq \mathsf{g}_*(T)T^4$, where $\mathsf{g}_*(T)$ counts relativistic degrees of freedom at a given temperature and is slightly bigger than $100$ above the temperature $300$\,GeV in the framework of standard model of particle physics. One more point worth paying attention is that the energy density \eqref{adreuliskalveli} is close to $\rho_\nu^0$ even if the temperature is taken as high as $1$TeV. Recall that the kinetic term is suppressed by the huge parameter $\beta$, which is at least of the order of $10^{58}$. One will easily find that the kinetic term at earlier times is of the order of the potential one. What happens at later times is that the kinetic term decreases as $H$ gets smaller and the compound of scalar field and the CNB start to act as a dark energy after CNB becomes non-relativistic.

  Above the decoupling temperature one has to use the equilibrium distribution, which for the effective potential gives $U(\phi) - p_\nu(\phi, T_\nu)$. One can easily derive it by noting that in this case (see Eq.\eqref{tsonastsoruli})

\begin{eqnarray}
	&&\frac{m'_\nu}{m_\nu} (\rho_\nu - 3p_\nu) = \nonumber \\&&  \frac{ \mathsf{g} }{a^3}\frac{\mathrm{d}}{\mathrm{d}\phi} \int\frac{\mathrm{d}^3k}{(2\pi)^3}\, \left(\varepsilon_\nu(\mathbf{k}) - T_\nu \ln\left[1+\mathrm{e}^{\varepsilon_\nu(\mathbf{k})/T_\nu} \right] \right) ~, \nonumber 
\end{eqnarray} which after using an integration by parts gives

\begin{eqnarray}
 &&4\pi \int_0^\infty\frac{\mathrm{d}k k^2}{(2\pi)^3}\, \left(\varepsilon_\nu(\mathbf{k}) - T_\nu \ln\left[1+\mathrm{e}^{\varepsilon_\nu(\mathbf{k})/T_\nu} \right] \right) = \nonumber \\&&  - \frac{4\pi}{3} \int_0^\infty\frac{\mathrm{d}k k^3}{(2\pi)^3}\, \frac{\mathrm{d}}{\mathrm{d}k} \left(\varepsilon_\nu(\mathbf{k}) - T_\nu \ln\left[1+\mathrm{e}^{\varepsilon_\nu(\mathbf{k})/T_\nu} \right] \right) = \nonumber \\&& - \frac{1}{3} \int\frac{\mathrm{d}^3k }{(2\pi)^3}\, \frac{k^2}{\varepsilon_\nu(\mathbf{k})\left(1+\mathrm{e}^{\varepsilon_\nu(\mathbf{k})/T_\nu} \right) } ~. \nonumber 
\end{eqnarray} Thus, one finds

\begin{eqnarray}
&&\frac{m'_\nu}{m_\nu} (\rho_\nu - 3p_\nu) =  -\frac{\mathrm{d}p_\nu(\phi, T_\nu)}{\mathrm{d}\phi}  ~. \nonumber 
\end{eqnarray} Expanding $p_\nu(\phi, T_\nu)$ in a power series in $m_\nu^2/T_\nu^2$,

\begin{eqnarray}
	p_\nu(\phi, T_\nu) = \frac{\mathsf{g}T_\nu^4}{3}\left[\frac{7\pi^2}{240} -\frac{m_\nu^2}{16T_\nu^2} +O\left(\frac{m_\nu^4}{T_\nu^4}\right) \right] ~, \nonumber 
\end{eqnarray} and comparing it with Eq.\eqref{numbered}, one infers that the minima of $U_{eff}$ are again given by the Eq.\eqref{relatminimumebi}. Therefore, the consequent conclusions hold above the decoupling temperature as well.

The thermal equilibrium stage is preceded by the particle production in the post-inflation epoch. In the present model the conventional preheating mechanism \cite{Kofman:1994rk, Kofman:1997yn} can operate successfully for creating the cosmic fluid out of thermal equilibrium which then undergoes the thermalization. Instant preheating, which is inevitable for the runaway type potentials of quintessential inflation having no oscillation regime \cite{Geng:2015fla, Ahmad:2017itq, Geng:2017mic}, is not required in the present case. There is, however, a subtle point concerning the naturalness. The preheating is usually achieved by introducing a coupling of matter field with the inflaton that results in a time-dependent mass term. Loosely speaking, in certain regions of the parameter space, the solution of the matter field in this time-dependent background grows rapidly corresponding to what is called parametric resonance. One will find that the particle production within a broad resonance regime is big enough draining rapidly the energy from oscillating inflaton field. However, the coupling of $\phi$ with $\psi_\nu$ in the present model does not result in the neutrino production as it provides a real mass variation of neutrinos. It maybe somewhat unnatural that neutrinos represent exception to the general rule. This problem of naturalness persists for the instant preheating as well.

\section{Neutrino lumps}
\label{nnuggets}

Concerning the perturbations, most subtle and interesting issue is the possibility of formation of the neutrino clumps \cite{Afshordi:2005ym, Bjaelde:2007ki, Bean:2007ny, Bean:2007nx}. Instead of deriving instabilities via the effective sound speed of the compound of scalar field and neutrino fluid, we shall approach this problem from a somewhat different point of view. As the perturbations are of quantum origin, let us consider quantum fluctuations both for scalar and fermion fields. For this purpose, the scalar field is split as $\phi +\chi$ and the fermion number operator is shifted as $\bar{\psi}_\nu\psi_\nu \to n_\nu +  \bar{\psi}_\nu\psi_\nu$. Forgetting about the gravitation, the Lagrangian density for the perturbations takes the form  

\begin{eqnarray}\label{perturbationL} \frac{\partial_\alpha\chi\partial^\alpha\chi}{2} - \frac{\Big(U''(\phi_+)+m''_\nu(\phi_+)n_\nu\Big)\chi^2}{2}  + \nonumber \\ i\bar{\psi}\gamma^\alpha\partial_\alpha\psi \ - m_\nu'(\phi_+)\chi\bar{\psi}\psi + \text{C.C.} + \text{H.T.}~,  
\end{eqnarray} where C.C. denotes complex conjugate and H.T. stands for the higher order terms. From Eq.\eqref{perturbationL} one sees that there is an attractive force between the neutrinos mediated by the exchange of $\chi$ quanta. It results in the Yukawa potential

\begin{eqnarray}\label{Yukawa}
- \, \frac{\big( m_\nu'(\phi_+)\big)^2\exp\Big(-\sqrt{U''(\phi_+)+m''_\nu(\phi_+)n_\nu}\,r\Big)}{4\pi r} ~, ~~
\end{eqnarray} where $r$ stands for the physical distance, implying that the corresponding attractive force is characterized with the screening length

\begin{eqnarray}
\frac{1}{\sqrt{U''(\phi_+)+m''_\nu(\phi_+)n_\nu}} = \frac{1}{\sqrt{U_{eff}''(\phi_+)}} \equiv \frac{1}{m_{eff}(\phi_+)} ~. \nonumber 
\end{eqnarray} It is instructive to compare this force, within its screening radius, with the gravitational one. The ratio is

\begin{eqnarray}
	\big(m_\nu'(\phi_+)\big)^2 \, \frac{M_P^2}{m^2_\nu(\phi_+)} = 36\beta ~. \nonumber 
\end{eqnarray} One sees that the fifth force exceeds the Newtonian one by many orders of magnitude. As to the effective mass, it reads

\begin{eqnarray}\label{efeqturi}
	m_{eff}^2  \, \simeq \,  \frac{ 36\beta \rho_\nu}{M_P^2} \,\simeq\, \beta H_0^2  ~.  
\end{eqnarray}

Proceeding in the spirit of the above discussion, we treat the neutrino gas as an ideal pressureless fluid subject to the Newtonian self-gravity and also to the Yukawa force. In addition, we assume an expanding background, $\mathbf{r} =  a(t)\mathbf{x} $, to avoid the "Jeans swindle" \cite{Kiessling:1999eq}. That is the well known formalism one can find in many textbooks on cosmology, see for instance \cite{Peebles:1994xt}. The linear perturbations of neutrino velocity field and density contrast, $\delta \equiv \delta\rho_\nu/\rho_\nu$, satisfy the equations

\begin{eqnarray}\label{VF}
	&&\frac{\partial\delta}{\partial t} + \frac{\nabla_{\mathbf{x}}\cdot \delta \mathbf{v}}{a} = 0 ~,  \\&&\label{DC} \frac{\partial\delta\mathbf{v}}{\partial t} +H\delta\mathbf{v} + \frac{c_s^2\nabla_{\mathbf{x}}\delta}{a} + \nonumber \\ && \frac{\nabla_{\mathbf{x}}\Phi_N}{a}+ \frac{m'_\nu(\phi_+)}{m_\nu(\phi_+)} \frac{\nabla_{\mathbf{x}}\Phi_Y}{a} = 0 ~,  ~~
\end{eqnarray} where $\Phi_{N,Y}$ denote the Newton and Yukawa potentials, respectively:

\begin{eqnarray}&&\label{NP}
	\frac{\Delta_{\mathbf{x}}\Phi_N}{a^2} = \frac{4\pi}{M_P^2} \, \rho_\nu\delta ~,  \\&&\label{YP}  \left(\frac{\Delta_{\mathbf{x}} }{a^2} - m_{eff}^2(\phi_+)\right) \Phi_Y = \frac{m'_\nu(\phi_+)}{m_\nu(\phi_+)} \, \rho_\nu\delta ~. 
\end{eqnarray} First we take the divergence of Eq.\eqref{DC} and substitute in it $\nabla_{\mathbf{x}}\cdot \delta \mathbf{v}$ from Eq.\eqref{VF} 

\begin{eqnarray}\label{PEQ}
&&	\frac{\partial^2\delta}{\partial t^2} + 2H \frac{\partial\delta}{\partial t} - \nonumber \\&& \frac{c_s^2\Delta_{\mathbf{x}}\delta}{a^2} - \frac{\Delta_{\mathbf{x}}\big(\Phi_N + \Phi_Y m'_\nu(\phi_+)/m_\nu(\phi_+)\big)}{a^2}   =0 ~.  ~~~~
\end{eqnarray} Applying now Fourier decomposition for the density contrast and using Fourier transform of $\Phi_{N,Y}$ from Eqs.(\ref{NP}, \ref{YP}), the Eq.\eqref{PEQ} takes the form

\begin{eqnarray}\label{tsrpivi}
&&\ddot{\delta}(\mathbf{k}) +2H\dot{\delta}(\mathbf{k}) +  \left(\frac{c_s^2k^2}{a^2} \,-\,  \frac{4\pi\rho_\nu }{M_P^2}  \,-\, \right. \nonumber \\&& \left. \left(\frac{m'_\nu(\phi_+)}{m_\nu(\phi_+)}\right)^2 \frac{\rho_\nu k^2/a^2}{k^2/a^2+m_{eff}^2(\phi_+)}\right)\delta(\mathbf{k})  = 0 ~. 
\end{eqnarray} Much of the essential physics concerning the instabilities of neutrino perturbations can be extracted from this equation. One immediately sees that the Yukawa force helps the amplification of density perturbations and, exploiting the idea similar to what was suggested in \cite{Bjaelde:2007ki}, one can incorporate the scalar-mediated force with the gravitational one by introducing an effective Planck mass 

\begin{eqnarray}
	\frac{1}{M^2_{eff}} \equiv \frac{4\pi }{M_P^2} +  \left(\frac{m'_\nu(\phi_+)}{m_\nu(\phi_+)}\right)^2 \frac{k^2/a^2}{k^2/a^2+m_{eff}^2(\phi_+)} \simeq \nonumber \\ \frac{1}{M_P^2}\left(4\pi + \frac{\beta k^2/a^2}{k^2/a^2+m_{eff}^2(\phi_+)} \right) ~. ~~~~~~~ \nonumber 
\end{eqnarray} From this expression one sees that for $k\gg m_{eff}$ the Newton constant is amplified at least by the factor of the order of $10^{58}$ leading to the growth of small neutrino perturbations. Without the Yukawa amplification, there would be no growing perturbations of reasonable size (see below). Assuming for simplicity that $\rho_\nu, a, c_s$ are constants, one finds the solution of Eq.\eqref{tsrpivi} 

\begin{eqnarray}
	\delta(t, \mathbf{k}) = C_1(\mathbf{k})\exp \left(t\sqrt{\frac{\rho_\nu}{M^2_{eff}} - \frac{c_s^2k^2}{a^2}}\right) + \nonumber \\ C_2(\mathbf{k})\exp \left(-t\sqrt{\frac{\rho_\nu}{M^2_{eff}} - \frac{c_s^2k^2}{a^2}}\right) ~, \nonumber 
\end{eqnarray} manifesting the possibility that even for vanishing gravity one may have the sub-horizon modes growing at the expense of Yukawa force (here we use Eq.\eqref{efeqturi})

\begin{eqnarray}
 H_0^2 <  \frac{k^2}{a^2} \,<\, \frac{\rho_\nu}{c_s^2} \left(\frac{m'_\nu(\phi_+)}{m_\nu(\phi_+)}\right)^2 - m^2_{eff} \simeq ~~~~~~ \nonumber \\  \left(\frac{1}{c_s^2} -1\right) m^2_{eff}   = \left(\frac{1}{c_s^2} -1\right)\beta H_0^2  ~. \nonumber 
\end{eqnarray} Behind this expression, one easily recognizes similar estimates from the previous works \cite{Bean:2007ny, Bean:2007nx}. The only difference is that now it is augmented by the factor $c_s^{-2}-1$. The speed of sound for the non-relativistic matter estimated as \cite{Peebles:1987ek} $c_s \sim \sqrt{T_\nu/m_\nu}$ makes it easy to see why the formation of nuggets becomes favorable in the non-relativistic regime. In general, the growing modes, with respect to Eq.\eqref{tsrpivi}, satisfy the condition 

\begin{widetext}
\begin{eqnarray}
&&	\frac{k^2}{a^2} \,<\, \left\{  \frac{4\pi\rho_\nu }{M_P^2} + \left(\frac{m'_\nu(\phi_+)}{m_\nu(\phi_+)}\right)^2 \rho_\nu - \frac{c_s^2m_{eff}^2(\phi_+)}{a^2} + \sqrt{\mathfrak{D}} \right\}\frac{1}{2c_s^2} ~,  \nonumber \\&& \text{where} ~~\mathfrak{D} = \left( \frac{c_s^2m_{eff}^2(\phi_+)}{a^2} -  \frac{4\pi\rho_\nu }{M_P^2} - \left(\frac{m'_\nu(\phi_+)}{m_\nu(\phi_+)}\right)^2 \rho_\nu  \right)^2 + \frac{16\pi \rho_\nu m^2_{eff}c_s^2}{a^2M_P^2} ~. \nonumber 
\end{eqnarray}\end{widetext} In absence of the fifth force, the Jeans length would be 

\begin{eqnarray}
\frac{a}{k} \, \gtrsim \frac{11 c_s}{H_0} ~, \nonumber 
\end{eqnarray} manifesting the need of the fifth force for creating neutrino nuggets.

The above discussion is intended, on the one hand, to elucidate the qualitative features in a simple and transparent way and, on the other hand, our results form a new foundation for future investigations of formation and subsequent implications of nuggets. Let us note that the lumpy CNB is one of the direct observational consequences of the model and there are many papers devoted to the study of nugget formation process involving the non-linear dynamics \cite{Brouzakis:2007aq, Schrempp:2009kn, 2010PhRvD..81f3525W, 2010PhRvD..82j3516W, Savastano:2019zpr, Casas:2016duf, 2013PhRvD..87d3519A, 2012PhRvD..85l3010A}. The new feature emphasized in the above discussion is the appearance of effective mass in Eq.\eqref{YP} indicating the screened nature of the fifth-force. For the sake of comparison, see Eqs.(13-16) in \cite{2010PhRvD..81f3525W} and Eqs.(21-25) in \cite{2010PhRvD..82j3516W}. Further research can be conducted in the light of recent investigations of the Yukawa nuggets \cite{Wise:2014jva, Wise:2014ola, Gresham:2017zqi, Gresham:2018anj}.

\section{Discussion and conclusions}

 The idea behind the introduction of a non-standard coupling between the quintessence field and the CNB is to tie the dark energy density to the neutrino mass scale \cite{Fardon:2003eh}, see for instance Eq.\eqref{adreuliskalveli}. There are, however, two categories of such models. The first category of models known as "growing neutrino quintessence" \cite{Wetterich:2007kr, Amendola:2007yx} assume that the scalar field is steadily rolling down the potential $U(\phi)$ before the neutrinos become non-relativistic and stop its motion resulting in the potential-dominated dynamics for the quintessence field. We are interested in other category of models \cite{Pietroni:2005pv, MohseniSadjadi:2017pne} assuming that the scalar field is initially trapped into the vacuum and acquires a non-zero vacuum expectation value, which then varies in time adiabatically, as a result of back-reaction due to neutrinos. The latter scenario allows one to naturally incorporate well established inflationary models into it. Most successful inflationary models are believed to be those with a plateau \cite{Kallosh:2019hzo, Carrasco:2015rva}. Such models are characterized by two independent parameters - the width and the height of the potential. The example considered by us is a typical representative of such a model. It is characterized by the parameters  $\alpha$ and $V$, see section \ref{parameters}, where the dimensionless parameter $\alpha$ is of the order of unity. Besides, we have one more dimensionless parameter, $\beta \gtrsim 10^{58}$, which comes from the coupling term. We see a large discrepancy between these dimensionless parameters that should be "explained" somehow. Interestingly enough, the broad class of inflationary potentials derived in \cite{Kallosh:2013hoa} as a result of spontaneously broken conformal symmetry, can be straightforwardly used in the above discussion with the same $\phi$-$\nu$ coupling term (which is certainly taken by hand). Namely, the T-model potentials \cite{Carrasco:2015rva}

 \begin{eqnarray}
 U\left(\tanh^2 \frac{\phi}{\sqrt{6\alpha M_P^2}}\right)  ~, \nonumber 
 \end{eqnarray} are closely analogous to what we have considered and, therefore, it is straightforward to generalize our discussion to this kind of models. Also, it is almost obvious that the construction similar to what we have discussed may work without demanding $\mathcal{Z}_2$ symmetry for the inflaton potential. For instance, one may consider a Starobinsky-like model \cite{Starobinsky:1980te, Starobinsky:1983zz}, which would give 
 
 \begin{eqnarray}
 U_{eff} = V\Big(1-\mathrm{e}^{-\alpha \phi/M_P}\Big)^2 + n_\nu \mu_\nu\mathrm{e}^{-\beta\phi^2/M_P^2} ~. \nonumber
 \end{eqnarray} From the very outset one can make a simplifying approximation $\alpha \phi_+/M_P \ll 1$, which is well justified as long as $\beta \gg \alpha$. So that, for finding the minimums one can use an approximate expression   
 
 \begin{eqnarray}
 U_{eff} \approx \frac{V\alpha^2 \phi^2}{M^2_P} + n_\nu \mu_\nu\mathrm{e}^{-\beta\phi^2/M_P^2}  \Rightarrow   \frac{\phi_\pm}{M_P} \approx \sqrt{\frac{1}{\beta}\ln\frac{\beta n_\nu\mu_\nu}{\alpha^2 V}} ~. ~~~~~~~~~ \nonumber
 \end{eqnarray} Hence, one finds

 \begin{eqnarray}\label{staroneutr}
 m_\nu = \frac{\alpha^2 V}{\beta n_\nu} ~~\Rightarrow~~ \rho^0_\nu = \frac{\alpha^2V}{\beta} ~. 
 \end{eqnarray} The slow roll condition now takes the form

 \begin{eqnarray}\label{staroslow}
 \frac{9M_P^2H^2}{8} \ll 
 \alpha^2 V \ln^2\frac{\mu_\nu}{m_\nu(\phi_+)} ~, \nonumber 
 \end{eqnarray} and is satisfied with a high accuracy at present since in the Starobinsky model: $V\simeq 1.5\times 10^{-13}M^4_P, \alpha = \sqrt{2/3}\approx 0.8$ and, as we have already estimated: $\ln\mu_\nu/m_\nu(\phi_+)=9$. The value of $\beta$ can be estimated crudely from the Eq.\eqref{staroneutr} by noting that $\rho^0_\nu \lesssim \rho_c \simeq M_P^2H_0^2$ and correspondingly 
 
 \begin{eqnarray}
 \beta \gtrsim  \frac{V}{M_P^2H_0^2} ~. \nonumber 
 \end{eqnarray} We shall not continue this discussion as it precisely parallels what is already done in the text. Along the same lines of reasoning, one can construct E-type inflationary quintessence models given by the potential \cite{Carrasco:2015rva}
 
 \begin{eqnarray}
 U(\phi) = V\left(1-\exp\left(-\frac{\alpha\phi}{M_P}\right)\right)^{2n} ~. \nonumber 
 \end{eqnarray} 
 
 It is worth noting that the Starobinsky model is one of the most naturally motivated inflationary model, providing a good fit to the current observational data \cite{Aghanim:2018eyx, Akrami:2018odb}, that maybe considered as one of the target models for future observations \cite{Kallosh:2019hzo}.

In the framework of the present approach, we have derived the fifth-force as an Yukawa one arising due to exchange of a single $\chi$ boson between the neutrinos. This force gets corrected by the exchange of a pair of $\chi$ bosons

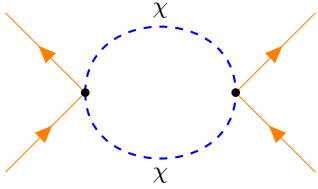
\begin{figure}[h]
	\centering
	\begin{tikzpicture}[baseline=(current bounding box.center)]
	\begin{feynman}
	\vertex (x);
	\vertex[right=2cm of x] (y);
	\vertex[above left=of x] (a);
	\vertex[below left=of x] (b);
	\vertex[above right=of y] (c);
	\vertex[below right=of y] (d);
	\vertex[above right= 0.5cm and 0.75cm of x] (p);
	
	\diagram*{
		(x) --[scalar, blue, thick, half right, edge label'=\(\color{black}{\chi}\)] (y),
		(x) --[scalar, blue, thick, half left, edge label=\(\color{black}{\chi}\)] (y),
		(x) --[fermion, orange] (a) ,
		(b) --[fermion, orange] (x),
		(y) --[fermion, orange] (c),
		(d) --[fermion, orange] (y),
	}; \draw[fill=black] (x) circle(0.5mm); \draw[fill=black] (y) circle(0.5mm);
	\end{feynman}
	\end{tikzpicture}\end{figure}

\noindent as long as quadratic perturbations are taken into account. It corresponds to the coupling 

\begin{eqnarray}
- \frac{m_\nu''(\phi_+)\chi^2\bar{\psi}\psi}{2} ~, \nonumber 
\end{eqnarray} and results in the potential \cite{Ferrer:1998rw} 

\begin{eqnarray}\label{fifth-force}
- \, \frac{\big( m_\nu''(\phi_+)\big)^2m_{eff}(\phi_+)K_1\Big(2m_{eff}(\phi_+)r\Big)}{32\pi^3 r^2} ~, \nonumber 
\end{eqnarray} where $K_1$ is the modified Bessel function. We have again a potential providing an attractive screened force, which for $r \lesssim m^{-1}_{eff} $ behaves roughly as 
 \begin{eqnarray}
- \, \frac{\big( m_\nu''(\phi_+)\big)^2}{64\pi^3 r^3}  ~ . \nonumber 
\end{eqnarray} This force starts to dominate over the Yukawa one \eqref{Yukawa} at relatively short distances

 \begin{eqnarray}
 	r \,\lesssim \,  \frac{\sqrt{\beta}}{M_P} \, \simeq \, 10^{-5}\text{cm} ~.  \nonumber 
 \end{eqnarray} Let us note that the length scale $\sqrt{\beta}/M_P$ is smaller by a factor of $10^{-5}$ than the range of Yukawa force: $1/m_{eff}$. Namely, for $\beta \simeq 10^{-58}$, one finds that (see Eq.\eqref{efeqturi})

 \begin{eqnarray}
 	\frac{1}{m_{eff}} \,\simeq\, 1\text{cm} ~. \nonumber 
 \end{eqnarray} This is a clear indication that for describing of nuggets one can safely use the Yukawa force without corrections unless the inter-neutrino distance is smaller than $10^{-5}$cm. However, by increasing the inflation energy scale, the $\beta$ parameter grows and this distance scale gets larger.

   One more aspect worth paying attention is that the force between the neutrinos mediated by the scalar field gets corrections if one assumes a finite temperature theory for the scalar field fluctuations \cite{Ferrer:1998rw}. That is to assume a thermal equilibrium between the $\chi$ field and neutrinos maintained by the $\chi$-$\nu$ coupling. So far there are very few papers addressing this issue. At least what we are familiar with are the papers \cite{Chitov:2009ph, Mandal:2019kkv} where the approach to the dark energy is somewhat different from the conceptual point of view but the techniques developed there can be readily used for the problem posed above.

\begin{acknowledgments} Authors are indebted to Tina Kahniashvili for her encouragement and helpful comments. The work was supported in part by the Rustaveli National Science Foundation of Georgia under Grant No. FR-19-8306.
\end{acknowledgments}

\nocite{*}

\bibliographystyle{apsrev4-1}
\bibliography{Literaturverzeichnis}

\end{document}